**ARTICLE** OPEN

# On-chip transverse-mode entangled photon pair source

Lan-Tian Feng[1,2], Ming Zhang[3], Xiao Xiong[1,2], Yang Chen[1,2], Hao Wu[3], Ming Li[3], Guo-Ping Guo [1,2], Guang-Can Guo[1,2], Dao-Xin Dai[3] and Xi-Feng Ren [1,2]

Integrated entangled photon pair source is an essential resource for both fundamental investigations and practical applications of quantum information science. Currently there have been several types of entanglement, among which the transverse-mode entanglement is becoming attractive because of its unique advantages. Here, we report an on-chip transverse-mode entangled photon pair source via the spontaneous four-wave mixing processes in a multimode silicon waveguide. Transverse-mode photon pairs are verified over multiple frequency channels within a bandwidth of ~2 THz, and a maximally entangled Bell state is also produced with a net fidelity of 0.96 ± 0.01. Our entangled photon pair source is the key element for quantum photonics based on transverse-mode, and also has the possibility to extend to higher-dimensional Hilbert space. Furthermore, the transverse-mode entanglement can be converted coherently to path and polarization entanglement, which paves the way to realizing highly complex quantum photonic circuits with multiple degrees of freedom.

*npj Quantum Information* (2019)5:2 ; https://doi.org/10.1038/s41534-018-0121-z

## INTRODUCTION

Quantum photon pair sources are not only critical to advancing our fundamental understanding of quantum mechanics, but also play a key role in many applications for quantum technologies.[1] Compared with free-space ones, integrated photon pair sources have attracted much attention owing to their compactness, scalability, and stability.[2] For example, path or polarization entangled photon pairs were demonstrated by using the non-linear processes in a single-mode waveguide or in a micro-ring cavity.[3–7] Aside from these degrees of freedom, transverse-mode in a multimode optical waveguide becomes more and more attractive, and shows advantages in modulation of wavelength, dispersion and polarization compared to single-mode ones.[8–11] It is useful for parallel encoding and enables information processing in a more compact way for next-generation optical communications.[12–14] On the other hand, via the nonlinear optical medium, photons in different orthogonal transverse-modes can interact with each other, which is useful for wavelength conversion and opens the door to many more nonlinear phenomena.[15–19]

Recently, transverse-mode has been applied for quantum photonics, for example, implementing quantum interference in one multimode waveguide and on-chip entanglement conversion between path, transverse-mode and polarization.[20–22] Also several works have demonstrated the photon pair generation with multiple modes in few mode fiber,[23] as well as in periodically poled KTiOPO$_4$ multimode waveguide.[24] These transverse-mode photon pairs in one multimode waveguide can be extended to higher-dimensional entangled states,[25] and therefore benefit highly-efficient quantum logic gates and noise-resilient communications.[26,27] Besides, transverse-mode entanglement can be converted coherently to path and polarization entanglement,[21] which showed the possibility to control multiple degrees of freedom of photons simultaneously and to realize chip-scale hyper-entangled states.[28] So far, on-chip transverse-mode entangled photon pair source has not yet been demonstrated.

Here, we study the spontaneous four-wave mixing (SFWM) processes in a multimode silicon nanophotonic waveguide thoroughly, and realize for the first time an on-chip transverse-mode entangled photon pair source. Transverse-mode entangled photon pairs are generated and experimentally verified with a bandwidth of ~2 THz. By adjusting the energy ratio and phase difference between different transverse-mode terms of the pump light, a maximally transverse-mode entangled Bell state is produced with a net fidelity of 0.96 ± 0.01. Our protocol can definitely be generalized to cases of higher-dimensional entangled photon pair sources with more transverse-modes, thus showing potential for realizing complex quantum states with high dimensionality and multiple degrees of freedom.

## RESULTS

### Entangled photon pair source chip

As shown in Fig. 1a, the device includes three parts: part I, the mode modulator; part II, the multimode waveguide; and part III, the output analyzer. The mode modulator is used to modulate the pump power ratio for different transverse-modes, and the output state analyzer is used to export the state for measurement. The multimode waveguide is 3 mm long, with a cross-section of ~760 × 220 nm², so that it can support two transverse-electric (TE) modes, i.e., TE$_0$ and TE$_1$. The pump light is coupled to the device through a grating coupler and then divided into two parts through a directional coupler, with one part being converted to the transverse-mode TE$_0$ and the other to the transverse-mode TE$_1$ through a mode multiplexer.[12] After nonlinear interactions in the multimode waveguide, a mode demultiplexer is used to separate the generated photons into different paths, which are then coupled out using two identical grating couplers for further measurement.

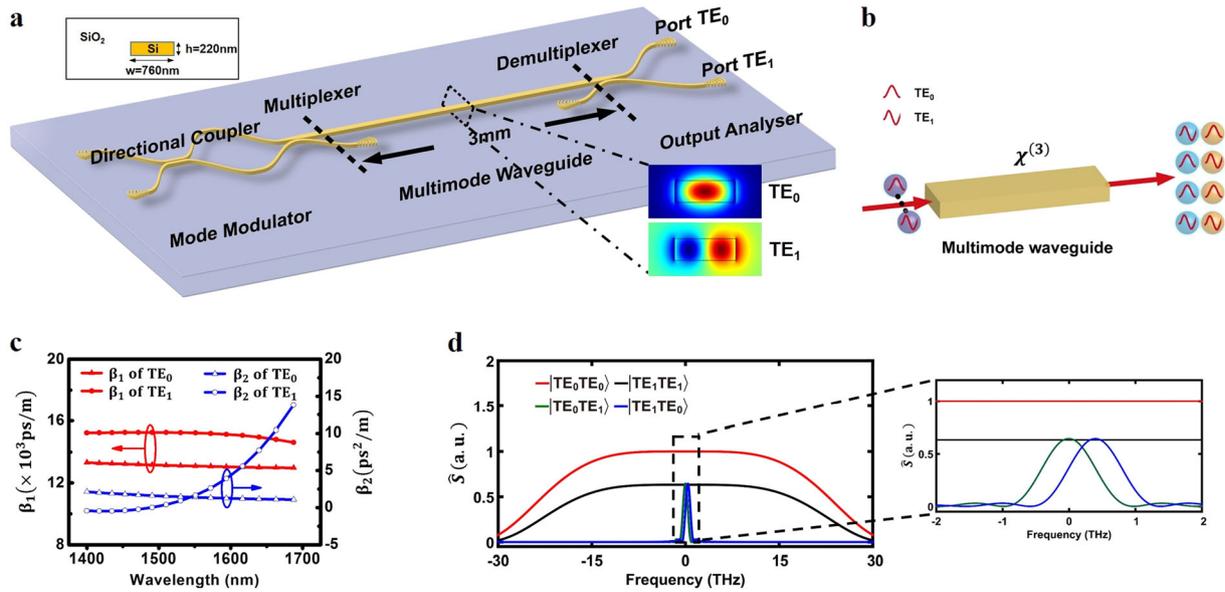

**Fig. 1** Schematic of the sample, concept of photon-pair generation and theoretical calculations. **a** Schematic of the sample, including a mode modulator, a multimode waveguide, and an output analyzer. Insets: cross-section of the multimode waveguide (upper left) and the $E_x$ components of the transverse-modes $TE_0$ and $TE_1$ (lower right). **b** Concept of photon-pair generation in a multimode waveguide. Pump photons can be in the same or different transverse-modes, and the generated idler and signal photons will be in multiple transverse-mode combinations. The SFWM processes in the multimode waveguide will generate correlated photon pairs in four kinds of transverse-mode combinations, i.e., $|TE_0TE_0\rangle$, $|TE_1TE_1\rangle$, $|TE_0TE_1\rangle$ and $|TE_1TE_0\rangle$. **c** The calculated value for the first-order and second-order dispersions of the multimode waveguide for $TE_0$ and $TE_1$ modes. **d** The calculated signal gain for four kinds of transverse-mode combinations of the generated photon pairs. All data are normalized to the maximum value. The curves for $|TE_0TE_1\rangle$ (green) and $|TE_1TE_0\rangle$ (blue) coincide, so we shifted the blue curve by 0.4 THz for a better view. Inset: zoom-in view

The SFWM processes in our device can be divided into three types (see Supplementary Materials, Section 1). For Type I, the four photons involved are in the same transverse-mode. For Type II, the two pump photons are in the same transverse-mode, while the signal and idler photons are in the other transverse-mode. For Type III, the two pump photons are in different transverse-modes (i.e., one in $TE_0$ and the other in $TE_1$), and the signal and idler photons are also in different transverse-modes. The SFWM process of Type I is intramodal, while the SFWM processes of Type II and Type III are intermodal. Here, the phase mismatching for the Type II process is so large that it hardly occurs. Therefore, we focus on Type I and Type III processes, where the signal and idler pairs have four combinations of transverse-modes, i.e., $|TE_0TE_0\rangle$, $|TE_1TE_1\rangle$, $|TE_0TE_1\rangle$ and $|TE_1TE_0\rangle$ (Fig. 1b). We calculated the first-order dispersions $\beta_1$ and second-order dispersions $\beta_2$ of the multimode waveguide for the $TE_0$ and $TE_1$ modes, respectively (Fig. 1c). The signal gain $\hat{S}$ of the generated photon pairs for four transverse-mode combinations is shown in Fig. 1d. All data are normalized to the maximum value. The signal and idler photons are time correlated, and their frequencies are equally separated from the central pump frequency.

**Photon pair source verification**

According to the size of waveguide at different parts, we calculated the SFWM efficiencies within parts I, II, and III, and concluded that the detected photon pair counts mainly came from the multimode waveguide part (see Materials and Methods). To ascertain the generated quantum states, we measured the photon pair coincidence for different combinations of the frequency and transverse-mode. The experimental setup is shown in Fig. 2a, an amplified continuous-wave (CW) laser (central wavelength 1550.11 nm) was coupled to the device via a grating coupler by a single-mode fiber array after passing through two cascaded 100 GHz bandwidth pre-filters. At the output end, the pump light was blocked by two cascaded off-chip 200 GHz bandwidth post-filters. Two 40-channel dense-wave-division-multiplexers (DWDMs) were used to separate the signal and idler photons. Each channel has a 100 GHz bandwidth, such that we can select any frequency- and transverse-mode combinations for photon pairs with a frequency detuning of ~2 THz from the central pump frequency (see Supplementary Materials, Table S3). The generated photon pairs were finally detected by two superconducting nanowire single photon detectors (SCONTEL, dark count rate 100 Hz, detection efficiency 85%).

We measured single photon counts in different frequency channels, as shown in Fig. 2b, c. The differences in single photon count arise from the different gains for the SFWM processes and the uneven Raman scattering and losses for different DWDM channels. To further characterize the photon source in the multimode waveguide, we also measured the coincidence counts for the processes of Type I (Fig. 2d, e) and Type III (Fig. 2f, g) processes. All coincidences are the raw data without subtracting the background and accident counts. As seen, the coincidences for the Type I processes are almost constant but accompanied by oscillation and the coincidences for the Type III processes decrease as the frequency detuning increases, which agrees well with the theoretical calculations (curves in Fig. 2d–g). The minor discrepancy may originate from the correlation between intramodal and intermodal SFWM processes and the difference in propagation loss of different transverse-modes that we ignored in theoretical calculation. Note that the photon pairs were collected with a longer integral time for the Type III process due to its lower collection efficiency.

Coincidence-to-accidental ratio (CAR) and source brightness are two key parameters for characterizing the photon pair source. We measured CARs for correlated photon pairs in different transverse-mode combinations (see Supplementary Materials, Fig. S1). Intramodal (Type I) photon pairs show high CARs (generally higher than 200 for $|TE_0TE_0\rangle$ and 100 for $|TE_1TE_1\rangle$), while the CARs for the intermodal photon pairs (Type III) are not high (lower than 10 in most cases) because of the large phase mismatching. It is possible to obtain higher CARs by broadening the frequency spectra with a wider multimode waveguide, benefiting from a smaller difference in the first-order dispersion between signal and idler photons.

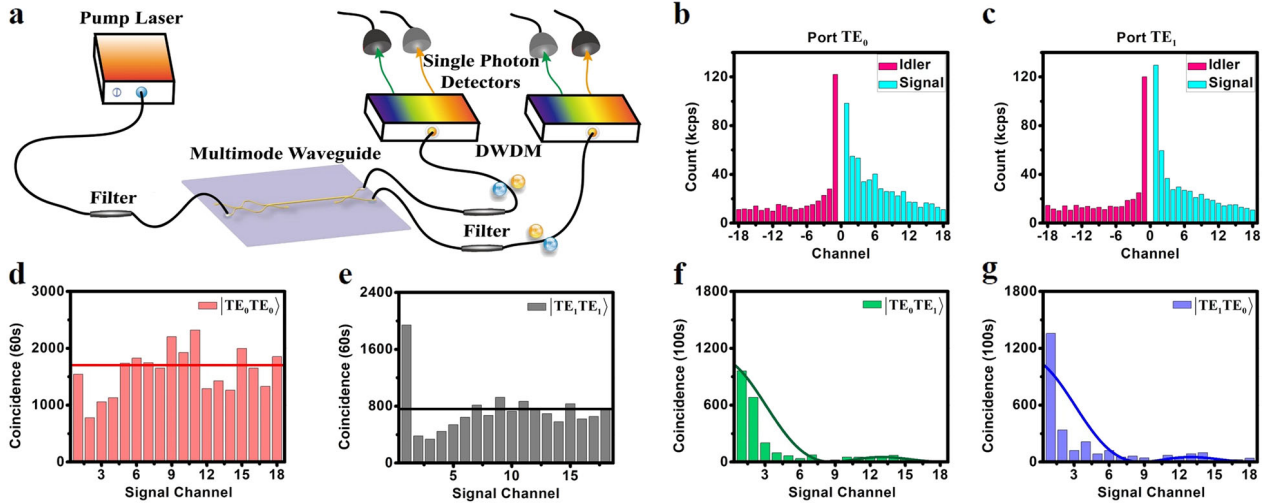

**Fig. 2** Sketch map for the experimental setup and measurement results. **a** An amplified CW laser is coupled to the device via a fiber array, and two DWDMs are used to separate the idler, signal and pump photons into different frequencies. **b, c** The continuous emission spectra for the signal and idler photons from ports TE₀ and TE₁, respectively. The spectra are filtered by a 40-channel DWDM system, with the frequencies of the correlated photon pairs equally separated from the central pump frequency. **d–g** Coincidence measurements of the photon pairs in different transverse-mode combinations. Solid curves are the corresponding theoretical results in Fig. 1d

In a multimode waveguide, the lower energy density enables a better tolerance to nonlinear noise; thus, one can simultaneously achieve high brightness and CARs by simply increasing the input pump power. In the experiment, we measured the intramodal photon pair $|TE_0TE_0\rangle$ ($|TE_1TE_1\rangle$) for channel ±8 with a calculated generation rate ranging from 19 kHz (5.5 kHz) to 530 kHz (180 kHz). The CARs did not suffer severe reduction even at the highest generation rates (see Supplementary Materials, Fig. S2). It also applies for the intermodal photon pair $|TE_0TE_1\rangle$ ($|TE_1TE_0\rangle$) for channel ±14, with a calculated generation rate ranging from 2.7 kHz (2.9 kHz) to 54 kHz (53 kHz) (see Supplementary Materials, Fig. S3). These results show that the present device provides a flexible platform for quantum information processing. It is worth pointing out that our multimode photon pair source could also be used as heralded single photon sources. Second-order correlation measurements show $g^{(2)}(0) = 0.13 \pm 0.02$ for photon pair $|TE_0TE_0\rangle$ and $g^{(2)}(0) = 0.19 \pm 0.06$ for photon pair $|TE_1TE_1\rangle$, respectively. More details are provided in the Supplementary Materials (Section 3 and Fig. S4).

In fact, silicon platform has been used for photon source generation with various structures.[29–31] A brief summary is provided in Supplementary Materials (Table S4). Compared to those single mode ones, the multimode waveguides have similar performance and could be further improved. Additionally, we also calculated the nonlinear efficiencies in a multimode waveguide as a function of the waveguide width (see Supplementary Materials, Fig. S5). For a waveguide with larger size, although the nonlinear efficiencies are decreased, a series of SFWM processes involving higher-order modes occur, and provides extra dimensions in quantum state generation.

## Entanglement verification and Bell state generation

After demonstrated the photon pair source generation, we then went further to demonstrate the generation of quantum entanglement, which lies at the heart of quantum information studies.[1] As described above, the quantum state in the multimode waveguide is given as

$$|\Phi\rangle = \frac{1}{\sqrt{1+\eta_1^2+\eta_2^2+\eta_3^2}} \left( |TE_1TE_1\rangle + \eta_1 e^{i\delta_1} |TE_1TE_0\rangle + \eta_2 e^{i\delta_2} |TE_0TE_1\rangle + \eta_3 e^{i\delta_3} |TE_0TE_0\rangle \right),$$ (1)

where $\eta_1$, $\eta_2$ and $\eta_3$ are relative proportions of different components; $\delta_1$, $\delta_2$ and $\delta_3$ are the phase differences, which come from the mode dispersions in the multimode waveguide. By using a two-dimensional (2D) grating coupler,[32,33] the transverse-mode entangled state is coherently converted into the polarization entangled state, which is expressed as $|\Phi\rangle = \frac{1}{\sqrt{1+\eta_1^2+\eta_2^2+\eta_3^2}}$

$\left( |HH\rangle + \eta_1 e^{i\delta_1} |HV\rangle + \eta_2 e^{i\delta_2} |VH\rangle + \eta_3 e^{i\delta_3} |VV\rangle \right).$

According to Figs. 1d and 2d–g, when the frequencies of the signal and idler photons are close to the pump frequency, all four transverse-mode combinations are generated. As an example, we performed the quantum state tomography measurement with the experimental setup shown in Fig. 3a for the photon pairs in the frequency channels ±2. Using $\{H, V, D, R\}^{\otimes 2}$ as the measuring basis, total 16 combinations are required to reconstruct the state density matrix, where $|H\rangle = (1, 0)^T$, $|V\rangle = (0, 1)^T$, $|D\rangle = \frac{1}{\sqrt{2}}(1, 1)^T$, $|R\rangle = \frac{1}{\sqrt{2}}(1, i)^T$ and $T$ is the transpose operator. Here we used the maximum-likelihood-estimation method[34] to reconstruct the density matrix $\hat{\rho}_{mea}$, as shown in Fig. 3b. Assuming the state is pure, the experimental data are fitted with the state expressed as $|\Phi\rangle = 0.60|TE_1TE_1\rangle + 0.29e^{-2.8i}|TE_1TE_0\rangle + 0.48e^{0.29i}|TE_0TE_1\rangle + 0.57e^{0.40i}|TE_0TE_0\rangle$). According to the definition $F = Tr(\hat{\rho}_{mea}\hat{\rho}_{pure})$, where Tr represents the trace and $\hat{\rho}_{pure} = |\Phi\rangle\langle\Phi|$, the net fidelity between these two density matrices is estimated to be $0.99 \pm 0.02$ (raw fidelity 0.93 ± 0.02). The high fidelity unambiguously shows that both intramodal and intermodal photon pairs are generated and that they constitute a complex but pure quantum state. The errors in fidelity were obtained by 100 times Monto Carlo calculation, with the experimental data subject to Gaussian statistics.

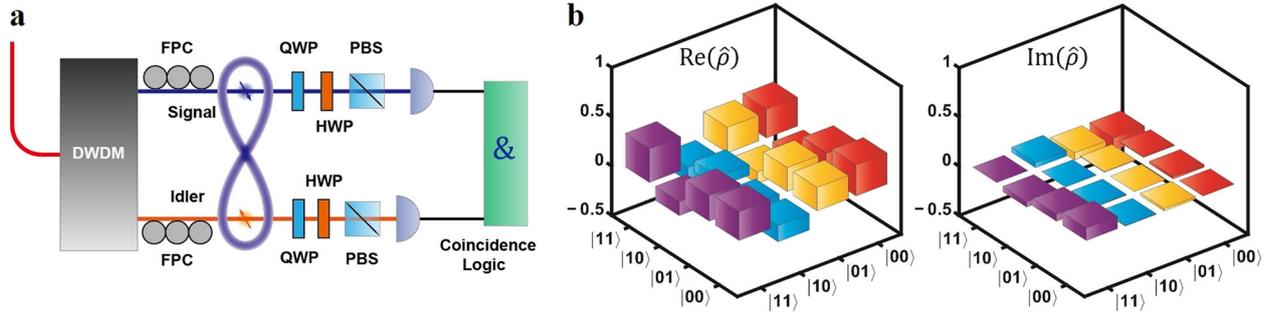

**Fig. 3** Quantum state characterization of photon pairs in the frequency channels ±2 by means of tomography. **a** Experimental setup for quantum state tomography. The polarizations of signal and idler photons exported from the DWDM are controlled by fiber polarization controllers (FPCs). The quarter-wave plate (QWP), half-wave plate (HWP) and polarization beam splitter (PBS) are tuned to measure coincidence between signal and idler photons and quantum state tomography. The photons are collected by two detectors and analyzed by the coincidence logic. **b** A quantum state can be fully described by its density matrix $\hat{\rho}$. The real (Re) and imaginary (Im) parts of the density matrix of the quantum state in the frequency channels ±2 were provided. Here, we use 1 to represent the transverse-mode $TE_1$ and 0 to represent the transverse-mode $TE_0$.

More interestingly, in some channels, intermodal photon pairs are negligible such that the entangled state

$$|\Phi\rangle = \frac{1}{\sqrt{1+\eta_3^2}}\left(|TE_1TE_1\rangle + \eta_3 e^{i\delta_3}|TE_0TE_0\rangle\right) \quad (2)$$

can be acquired. Furthermore, by adjusting the energy ratio and the phase difference between the $TE_0$ and $TE_1$ terms for the pump light, one can achieve a maximally transverse-mode entangled state as follows:

$$|\Phi\rangle = \frac{1}{\sqrt{2}}\left(|TE_1TE_1\rangle + |TE_0TE_0\rangle\right). \quad (3)$$

We took the frequency channels ±8 to test the quality of the entanglement. First, we remove the QWPs, and adjust the HWP in signal path to 0° or 45°, respectively. Then, we rotate the HWP in idler path from 0° to 360°. In this way, we obtained the coincidence between signal and idler paths as shown in Fig. 4a. The coincidence fringes can be fitted by $1 + V \sin[2\pi(\phi - \phi_c)/T]$, where $V$ is the fringe visibility, $\phi_c$ is the initial phase, and $T$ is the oscillation period. The fringe visibility $V$ is defined as $V = (d_{max} - d_{min})/(d_{max} + d_{min})$, where $d_{max}$ and $d_{min}$ are the maximum and minimum of the fitted data, respectively. The net visibilities for $\phi_s = 0°$ (solid black line) and $\phi_s = 45°$ (solid red line) bases are $97 \pm 2\%$ (raw visibility $95 \pm 2\%$) and $100 \pm 2\%$ (raw visibility $99 \pm 2\%$), which are both greater than $\frac{1}{\sqrt{2}} \approx 71\%$ and show the presence of Bell nonlocality between the signal and idler photons.

Then, quantum state tomography was performed to measure the state density matrix $\hat{\rho}$. The ideal density matrix of the maximally entangled state in Eq. (3) and the measured density matrix of the output states from frequency channels ±8 are shown in Fig. 4b, c, respectively. The fidelity is defined as $F = \text{Tr}(\hat{\rho}_{mea}\hat{\rho}_{ideal})$, where $\hat{\rho}_{ideal}$ is the ideal density matrix. We obtained a net fidelity of $0.96 \pm 0.01$ (raw fidelity $0.93 \pm 0.01$), confirming that the generated quantum state is of high quality and very close to the ideal maximally entangled state. The deviation of the fidelity from unity was mainly due to the errors in rotating the angles of the wave plates. Through use of a cascading on-chip Mach-Zehnder interferometer with thermal tuning to regulate the energy distribution and phase of different transverse-modes, any level of biphoton entanglement can be achieved.

## DISCUSSION

We have shown the photon pair and entangled Bell state generation based on transverse-mode in a multimode waveguide. Using larger waveguide that support more transverse-modes enables high-capacity information processing within a more compact chip than single-mode ones. Very recently, we have successfully demonstrated a 10-channel transverse-mode (de) multiplexer in a 2.3 μm-wide waveguide.[35] We believe that a higher-dimensional entangled state preparation will become feasible in the near future, even though several issues need to be addressed. For example, the efficiency of Type III SFWM processes have very narrow bandwidth and decreases sharply. And the dispersion of different transverse-modes also leads to the decoherence of the quantum superposed state. Nevertheless, for higher-dimensional Bell state generation, we just need to use Type I SFWM processes, and should avoid the cross terms induced by the Type III processes by selecting frequency channels. As for the dispersion, it can be compensated using delay line or by switching the transverse-modes with grating.[22] Taking a 4-dimensional quantum state preparation as an example, we discussed about the feasibility of realizing higher-dimensional entangled quantum state with one multimode waveguide (see Supplementary Materials, Section 4 and Figs. S6–S8). Also, we could hybridize the degrees of freedom of both path and transverse-mode for higher-dimensional quantum state generation. For example, a 16-dimensional spatial mode requires only 4 multimode waveguides that support 4 transverse-modes.

The fact that transverse-mode entangled state can be coherently converted into path and polarization entanglement provides convenience for large-scale quantum photonic integrated circuits (QPICs)[36–44] and hyper-entanglement generation.[28] Due to the non-uniform intermodal nonlinear process, photon pairs in different frequency channels are in different quantum states; thus, we can choose the quantum state as desired by selecting frequency channels, which is unimaginable for single-mode ones. This frequency-multiplexed transverse-mode entangled photon pair source in a multimode waveguide offers high selectivity and flexibility for realizing quantum applications. In principle, we can excite and measure arbitrary transverse-modes with the mode multiplexing technique, special quantum state also could be engineered with the intermodal nonlinear processes.

The intermodal SFWM process, where both the first-order and second-order dispersions can be engineered for phase matching, also can be used to realize frequency conversion between widely separated wavelengths. According to our calculations, in a multimode silicon waveguide with a cross-section of $\sim 1600 \times 220 \, nm^2$, conversion between wavelengths separated by 600 THz ($\sim 800$ nm) can be achieved (see Supplementary Materials, Fig. S9). On the other hand, the high quality cavities with multiple transverse-modes have also been explored to enhance the SFWM processes.[45–47]

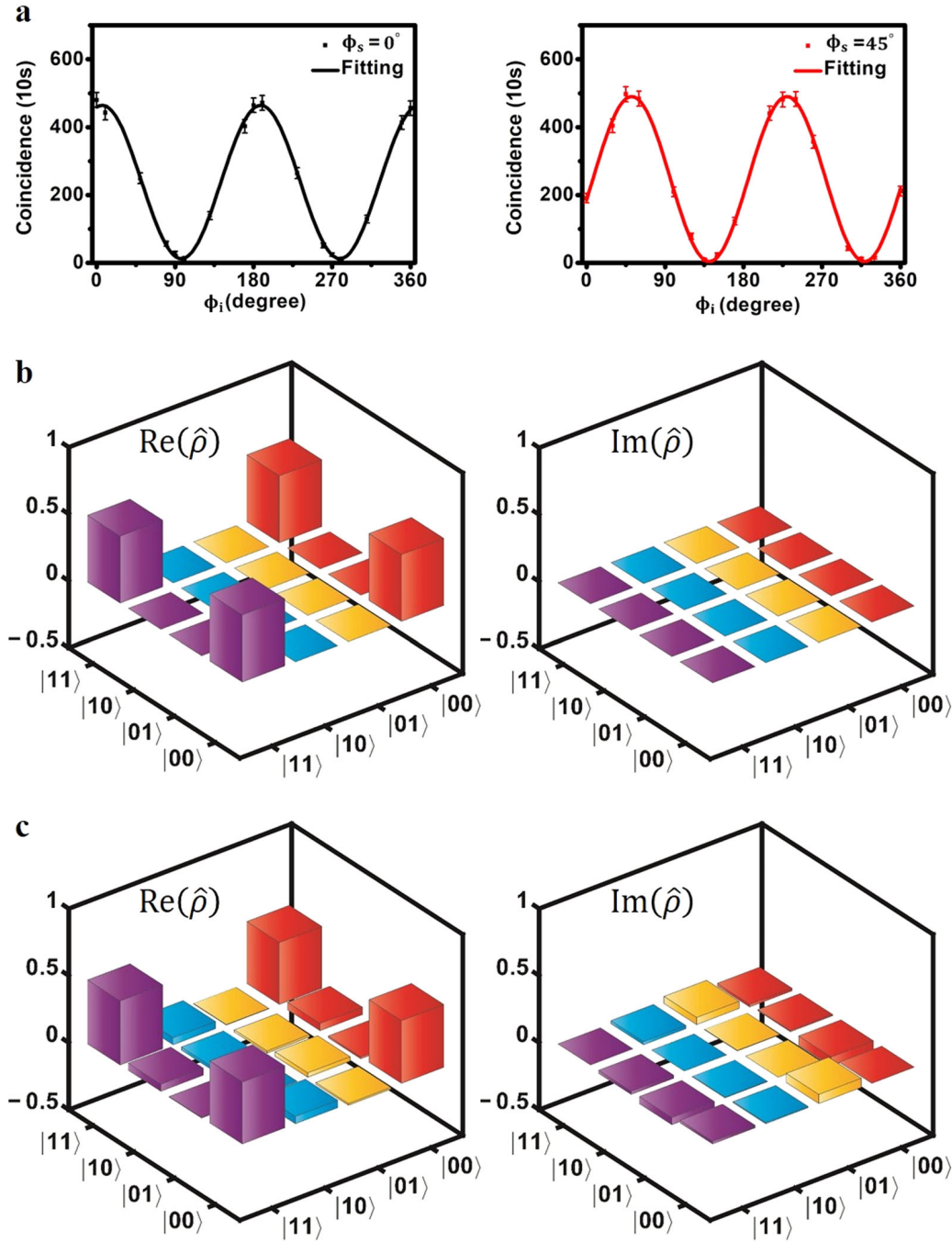

**Fig. 4** Quantum state characterization of photon pairs in the frequency channels ±8 through interference and state tomography. **a** Coincidences in 10 s intervals as a function of the idler polarizer angle when the signal polarizer is held at 0° (solid black line) and 45° (solid red line), respectively. **b** The real (Re) and imaginary (Im) parts for the density matrix of the ideal maximally entangled Bell state. **c** The measured density matrix of the quantum state agrees well with the ideal case. Here, we use 1 to represent the transverse-mode $TE_1$ and 0 to represent the transverse-mode $TE_0$

In conclusion, we have demonstrated the transverse-mode entangled photon pairs and quantum states generation using a multimode silicon waveguide, which has potential in extending further to higher-dimensional space. Combined with previous studies on coherent conversion and manipulation of transverse-mode entanglement,[21] the present work shows the potential for realizing complex quantum information processing with high dimensionality and multiple degrees of freedom. Moreover, the multimode waveguide enables to tailor quantum state with mode multiplexing technique and shows much richer nonlinear phenomena than the single-mode one, thus providing an attractive platform for quantum information applications.

## METHODS

### System efficiency

We ascertained the efficiencies for all components using laser light measurements (see Supplementary Materials, Fig. S10). The one-dimensional grating coupler and 2D grating coupler show coupling losses of 5 dB and 8 dB, respectively. The chip excess losses in our device are 5 dB for the TE$_0$ mode and 6 dB for the TE$_1$ mode. The mode multiplexer and demultiplexer induces an excess loss of 0.5 dB on the TE$_1$ mode. The post-filters and DWDM show an excess loss of 6 dB. The system used for state tomography shows a loss of 2.12 dB and both detectors have an efficiency of 85% ($-$0.7 dB).

### Photon pairs generated in different parts
Using the photon pair $|TE_0TE_0\rangle$ as example, we estimated the contribution of the finally detected photon pairs from parts I, II, and III. In our device, the part I and part III consist of single-mode waveguides with length of about 450 and 250 μm, respectively. Assuming that the pump power is the same and the conversion processes are lossless, the ratio of photon counts is $C_I$ : $C_{II}$ : $C_{III}$ = 1 : 188 : 0.1. When a transmission loss of 2 dB/mm is considered in the multimode waveguide (part II), the ratio became $C_I$ : $C_{II}$ : $C_{III}$ = 1 : 2.3 × 10$^4$ : 5.6 × 10$^{-8}$. Therefore, the photons generated in part I is mostly dissipated as propagation loss in part II. And in both cases, the contribution from single-mode waveguide (parts I and III) is negligible.

### Optical apparatus
We used a CW tunable laser (linewidth 10 kHz, central wavelength 1550.11 nm) as the pump light. The laser was amplified by an erbium-doped fiber amplifier and filtered (100 dB extinction ratio). The remaining power, which was coupled to the device after a single-mode fiber and an on-chip grating coupler, was 6.96 mW, which was then divided into two parts by an on-chip directional coupler to excite the two transverse-modes. Fiber alignment was maintained using a piezo-controlled four-dimensional displacement table for the adjustment of the position, as well as the coupling angle. The coupling angle was set as 15° and 4° for the one-dimensional grating coupler and 2D grating coupler, respectively. Two off-chip post-filters (100 dB extinction ratio) were used to remove the pump photons, and two DWDMs (with an extinction ratio of 30 dB for adjacent channels and 50 dB for non-adjacent channels) were used to separate the signal and idler photons. The correlated photons were recorded by two superconducting nanowire single photon detectors. The electrical signals were collected and analyzed through a time-correlated single photon counting (TCSPC) system, with the coincidence window set as 0.8 ns.

### Device fabrication
Chips were fabricated on the most commonly used 220 nm SOI platform. The pattern of the waveguides was exposed on ma-N 2403 negative tone photoresist using electron-beam lithography. The pattern was then transferred to the top silicon layer by an inductively coupled-plasma etching process. Grating couplers were fabricated using a second etching process to achieve efficient fiber-chip coupling. Finally, 1.2 μm plasma enhanced chemical vapor deposition (PECVD) SiO$_2$ was deposited on the top of the waveguide to form the upper-cladding.

### DATA AVAILABILITY
Data are available from the authors upon reasonable request.

### ACKNOWLEDGEMENTS
This work was supported by the National Natural Science Foundation of China (NSFC) (Nos. 61590932, 11774333, 61725503, 61431166001), the Anhui Initiative in Quantum Information Technologies (No. AHY130300), the Strategic Priority Research Program of the Chinese Academy of Sciences (No. XDB24030601), the National Key R & D Program (No. 2016YFA0301700), the Zhejiang Provincial Natural Science Foundation of China (Z18F050002), and the Fundamental Research Funds for the Central Universities. This work was partially carried out at the USTC Center for Micro and Nanoscale Research and Fabrication.

### AUTHOR CONTRIBUTIONS
All authors contributed extensively to the work presented in this paper. M.Z. and D.X. D. prepared the samples. L.T.F., M.Z., D.X.D., and X.F.R. performed the measurements, data analyses and discussions. X.X., Y.C., H.W., M.L., G.P.G., and G.C.G. conducted theoretical analysis. X.F.R. and D.X.D. wrote the manuscript and supervised the project.

### ADDITIONAL INFORMATION
**Supplementary information** accompanies the paper on the *npj Quantum Information* website (https://doi.org/10.1038/s41534-018-0121-z).

**Competing interests:** The authors declare no competing interests.

**Publisher's note:** Springer Nature remains neutral with regard to jurisdictional claims in published maps and institutional affiliations.